# Improving the efficiency of organic light emitting diodes by use of a diluted light-emitting layer

Siddharth Harikrishna Mohan [1], Kalyan Garre [2], Nikhil Bhandari [1], Marc Cahay [1]

[1] *Department of Electrical Engineering, University of Cincinnati, Cincinnati, Ohio 45220, USA*

[2] *Intel Corporation, 2501 NW 229th Ave, Hillsboro, Oregon 97124, USA*

## Abstract

The use of a thin mixed layer consisting of an inert diluent material and a light emitting material between the hole-transport layer and electron-transport layer of organic light-emitting diodes leads to an increase in the external quantum efficiency. The efficiency improvement is highly dependent on the thickness of the diluted light-emitting layer and driving current. Significant improvement seen at low current densities is explained in terms of effective hole confinement by the mixed layer while a modest decreases in efficiency at higher current densities may be attributed to luminescence quenching at the hole-transport layer/inert diluents material interface. The phenomena are demonstrated with three different inert diluents materials. A maximum external quantum efficiency improvement of about 40% is found for a diluted light-emitting layer thickness between 40 Å and 60 Å.

Organic light emitting diode (OLED) technology has generated much interest in the display and solid state lighting market owing to its superior features such as low operating voltage, wide color gamut, high efficiency and fast response. Since the report of first vacuum deposited bi-layer OLEDs by Tang and Van Slyke [1], tremendous progress has been made in improving their performance. Though the OLED field is rapidly growing, much time and effort must still be devoted to improving the OLED efficiency.

In most devices, the current is carried exclusively by holes in the hole-transport layer (HTL) and predominantly by electrons in the electron-transport layer (ETL). Simulation results by Tutis *et al.* [2] indicate a severe leakage of holes through the ETL to the cathode, particularly at low voltage, in an ITO/CuPc/NPB/Alq/LiF/Al structure, where ITO is indium tin oxide, CuPc is copper phthalocyanine, NPB is 4,4′-bis[*N*-(1-naphthyl)-*N*-phenylamino]biphenyl, and Alq is tris(quinolin-8-olato)aluminum(III). This leakage leads to a loss of efficiency of the OLEDs at low current densities. Similar devices without the CuPc layer also exhibit a relatively low external quantum efficiency (EQE) at low current densities [3, 4], and the reason may be the same. Alternatively, holes might not escape to the cathode, but recombine relatively close to the cathode where the resulting emission is outcoupled inefficiently. If so, the low efficiency could be improved by blocking the escape of holes or otherwise confining them within an emitting zone that is well placed for efficient optical outcoupling. The use of hole blocking layers (HBL) and quantum well structures has resulted in an improvement in the performance of undoped OLEDs [5-13]. Both these methods aim to improve the efficiency of OLEDs byimproving the utilization of all the injected carriers. Additional hole-blocking techniques such as insertion of a thin HBL of BCP (2,9-dimethyl-4,7-diphenylphenathroline) or Alq between a hole-injection layer (HIL) and the HTL [14] have been reported in the past. However, the inefficiency of NPB/Alq devices at low current densities has not specifically been addressed to date.

In this paper, we report the use of a thin mixed (co-deposited) layer of an inert diluent material (IDM) and Alq between an NPB HTL and an Alq ETL to increase the device EQE of an undoped OLED, especially at low current density. We individually investigate three IDMs in our mixed blocking layer structure and study the EQE improvement as a function of the layer thickness and the current density. Toward that goal, we used either HAT-CN or $CF_x$ for the HIL, NPB for the HTL, Alq for the ETL, BCP for an electron-injection layer, and 1:1 mixture of an IDM and Alq for the mixed EML. HAT-CN is *2,3,6,7,10,11*-hexacyano-*1,4,5,8,9,12*-hexaazatriphenylene and $CF_x$ is a plasma-polymerized fluorocarbon layer [15]. Three different non-emitting IDMs having very high solid-state ionization potentials (IP) were used: BCP, with an IP of 6.09 eV; hexaphenylbenzene (HPB), with an IP= 6.02 eV and 1,2,3,4-tetraphenylnapthalene (TPN), with an IP > 5.9 eV [16]. For comparison, the IP of Alq is 5.74 eV.[3] The OLEDs were fabricated on glass substrates pre-coated with a 25 nm layer of ITO as the anode. The substrates were scrubbed in a detergent solution, rinsed with deionized water, dried, and exposed to oxygen plasma for approximately 1 min [17]. $CF_x$ hole-injecting layer was deposited by plasma polymerization of $CHF_3$ on the ITO [15]. The substrates were then transferred into a vacuum chamber, and the organic layers were deposited sequentially by thermal evaporation under a vacuum of $10^{-6}$–$10^{-7}$ Torr, followed by the bilayer cathode consisting of 0.5 nm of LiF overlaid by 100 nm of aluminum.

The device structure was ITO/HAT-CN(100 Å, for BCP devices) or $CF_x$(10 Å, for TPN and HPB devices)/NPB(500 Å)/HBM:Alq(1:1, d Å)/Alq(400–d Å)/BCP(200 Å)/LiF(5 Å)/Al( 1000 Å). For instance, d = 20 Å would correspond to 10 Å of IDM co-deposited with 10 Å of Alq and so forth. The thickness d was varied in steps from 0 Å to a maximum of 80 Å. The neat BCP layer was included because preliminary experiments indicated that it improved the EQE somewhat. The total organic thickness was kept a constant to maintain the same optical outcoupling by adjusting the neat Alq layer. The entire device was hermetically encapsulated, together with a desiccant, under dry nitrogen. The EL properties of the OLEDs were evaluated using a Photoresearch PR650 SpectraScan Colorimeter and a programmable constant-current source meter at room temperature.

Figure 1 shows the EQE plotted versus current density (mA/cm$^2$) for different thicknesses of the TPN:Alq mixed EML. In the control device with an undiluted EML (NPB/Alq/BCP), the EQE falls off significantly at low current densities. This falloff has been observed before in simple NPB/Alq OLEDs [3, 4]. A 20 Å mixed layer increases the EQE by 25% at low current density (0.1 mA/cm$^2$), and by 10% at high current density (100 mA/cm$^2$). Similarly, a 30 Å mixed layer increases the EQE at all current densities but mainly remedies the low EQE at low current densities. Thicker layers (40 Å and above) show a consistent trend with significant efficiency improvement at low current densities. For example, the 60 Å device shows the best performance with EQE 1.1% at 0.1 mA/cm$^2$ which corresponds to an EQE improvement of around 40%. Though the thicker diluted light-emitting layers maintain the efficiency at low current densities, they tend to produce a falloff at high current densities. All of the EL spectra are similar to that of a simple NPB/Alq device, peaking at 526 – 528 nm [1]. There is a slight, progressive blue shift with increasing thickness of the mixed layer (up to 2 nm). No blue emission is observed from NPB, suggesting that this minor blue shift is solely due to the emission properties of diluted Alq. The spectra do not change with current density. Therefore, the luminous yield (cd/A) of all the devices shows the same trend as the EQE.

Preliminary measurements indicate an increased photoluminescence (PL) intensity in all IDM:Alq mixed films (TPN:Alq, HPB:Alq and BCP:Alq) when compared to undiluted Alq thin films. Although the quantum yield (QY) was not measured, it appears that the IDM somehow makes Alq more luminescent. In a similar finding, Fukuno *et al.* have observed a 25% increase of the PL QY in a 1:1 BCP:Alq film relative to a neat Alq film, which they attributed to a less polar environment in BCP:Alq [18]. An alternative explanation is that dilution of the Alq reduces self-quenching. They also observed a blue shift relative to neat Alq, consistent with the slight blue shift of the EL in our devices. They reported an increase of the EQE of an OLED using BCP:Alq as the light-emitting layer, but did not report on the current dependence [18]. While an increased QY of TPN:Alq may be a factor at all current densities, it does not readily explain the greater improvement at low current densities. If the control device suffers from hole leakage through the EML. Dilution may remedy the problem. If the leakage is most severe in the regime of low current densities, naturally the resulting efficiency improvement is largest there.

As TPN has a deeper HOMO than Alq, any conduction of holes in the mixed layer occurs mostly through the shallower HOMO levels of Alq molecule. Thus, TPN, the diluent in Alq, reduces the ability of the mixed layer to transport holes, impedes the escape of holes, and thereby increases the efficiency. Since the LUMO of TPN is higher than that of Alq [19], the ability to transport electrons is also reduced in the mixed layer. The result may be a build up of an electron space charge in the mixed layer, thus increasing the electric field at the HTL/EML interface, and requiring a greater density of positive charge at the interface. (The drive voltage might have been expected to increase, but the observed variation is small (<0.05 V) and not entirely consistent). The drive voltages were all about 3.6 V at 1 mA/cm$^2$ and 6.2 V at 20 mA/cm$^2$. It has been proposed that holes (NPB cation radicals) quench the emission of Alq [4]. A higher density of holes should produce stronger quenching, perhaps accounting for the decrease in efficiency at high current densities with the thicker diluted light-emitting layers.

Figures 2 and 3 show the EQE variation for devices implementing HPB:Alq and BCP:Alq mixed layers, respectively. The results are similar to those obtained with TPN:Alq. At a current density of 0.1 mA/cm$^2$, the 60 Å device with a HPB:Alq layer and the 40 Å device with the BCP:Alq layer shows their best performance with an EQE improvement of around 40% (EQE 1.12%)  The presence of the mixed layers again results in a slight blue shift of the emission. Effects on the drive voltage are small and nonsystematic. Both HPB and BCP have a large ionization potential similar to that of TPN, and the electron affinities of all three compounds are significantly smaller than that of Alq, and much smaller in the cases of TPN and HPB [16,19]. With these features in common, the similar results are consistent with a simple dilution effect. There are, however, differences in detail among the three diluents. From Figures 1 and 3, the EQE of the 20 Å TPN:Alq and BCP:Alq devices is higher than for the neat NPB/Alq device (0 Å) through the entire range of current densities. However from Fig. 2, the performance improvement of the 20 Å HPB:Alq device becomes marginal at high current densities. Thicker layers of BCP:Alq, TPN:Alq and HPB:Alq all produce a detrimental roll off in EQE at high current densities, but the thickness dependencies are somewhat different.

In conclusion, we have shown that the use of a thin mixed layer of an inert material (IDM) and Alq between an NPB HTL and an Alq ETL leads to an increase in the EQE of an OLED, especially when operated at low current densities. For ITO/CF$_x$/NPB/IDM:Alq/Alq/BCP/LiF/Al OLEDs, maximum EQE improvements of 40% were found with IDM:Alq layers consisting of TPN:Alq(1:1) and HPB:Alq(1:1), respectively. Similarly, for ITO/HAT-CN/NPB/IDM:Alq/Alq/BCP/LiF/Al OLEDs, a maximum EQE improvement of 40% was found with a IDM:Alq layer consisting of BCP:Alq(1:1), compared to devices without the IDM:Alq layer. The IDM:Alq layer thus introduces a new way to manipulate the charge transport and hence shape the recombination zone. Dilution apparently makes the EML more luminescent and remedies the leakage of hole though the EML at low current densities.

**Acknowledgment**

We thank Dr. Ralph Young for valuable suggestions during the OLEDs design, fabrication and characterization and for critical reading of the manuscript.

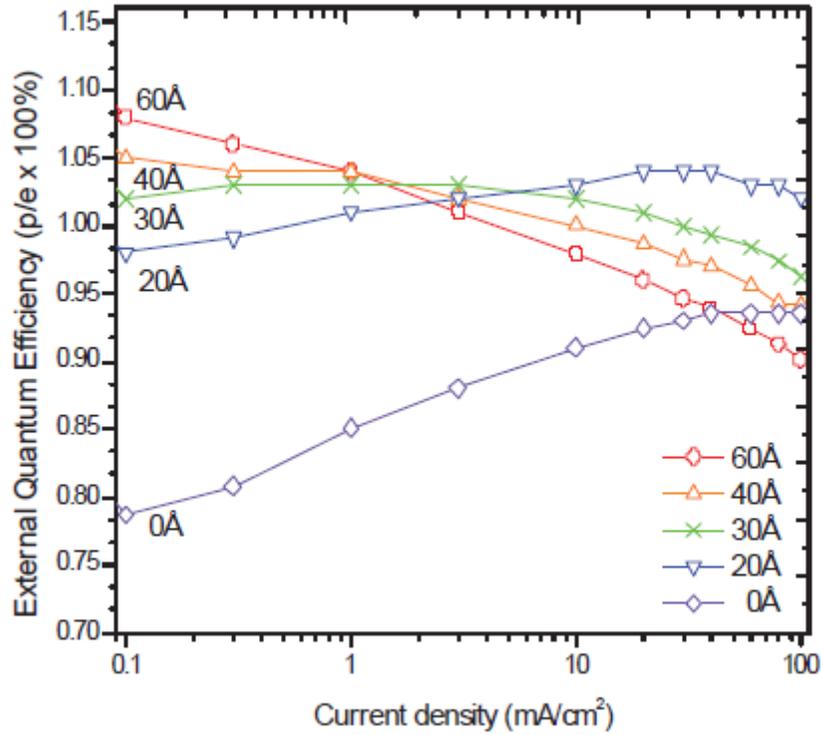

**Figure 1:** External quantum efficiency versus current density for different thicknesses d of the TPN:Alq diluted light-emitting layer in an ITO/CF$_x$/NPB/TPN:Alq(1:1,dÅ)/Alq/BCP/LiF/Al OLED. Also shown for comparison is the EQE of a reference OLED without the mixed layer (d = 0 Å).

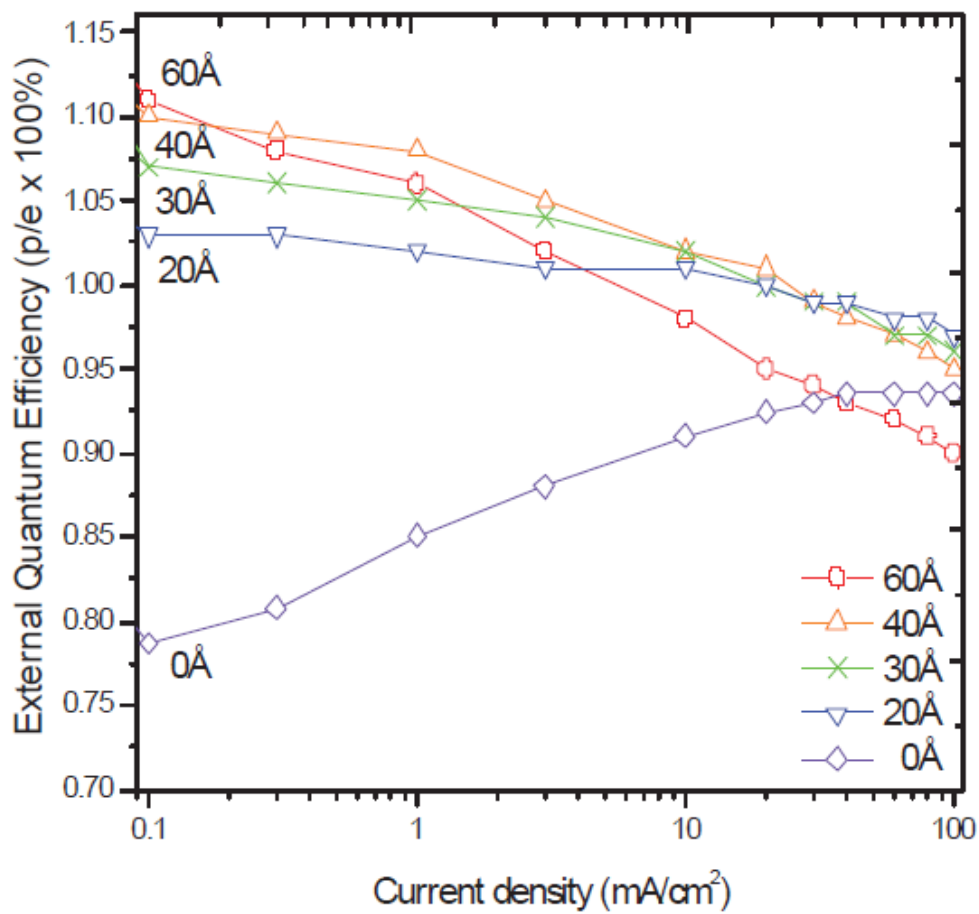

**Figure 2:** Same as Fig.1 but using a light-emitting layer diluted with HPB in an ITO/CF$_x$/NPB/HPB:Alq(1:1,dÅ)/Alq/BCP/LiF/Al OLED.

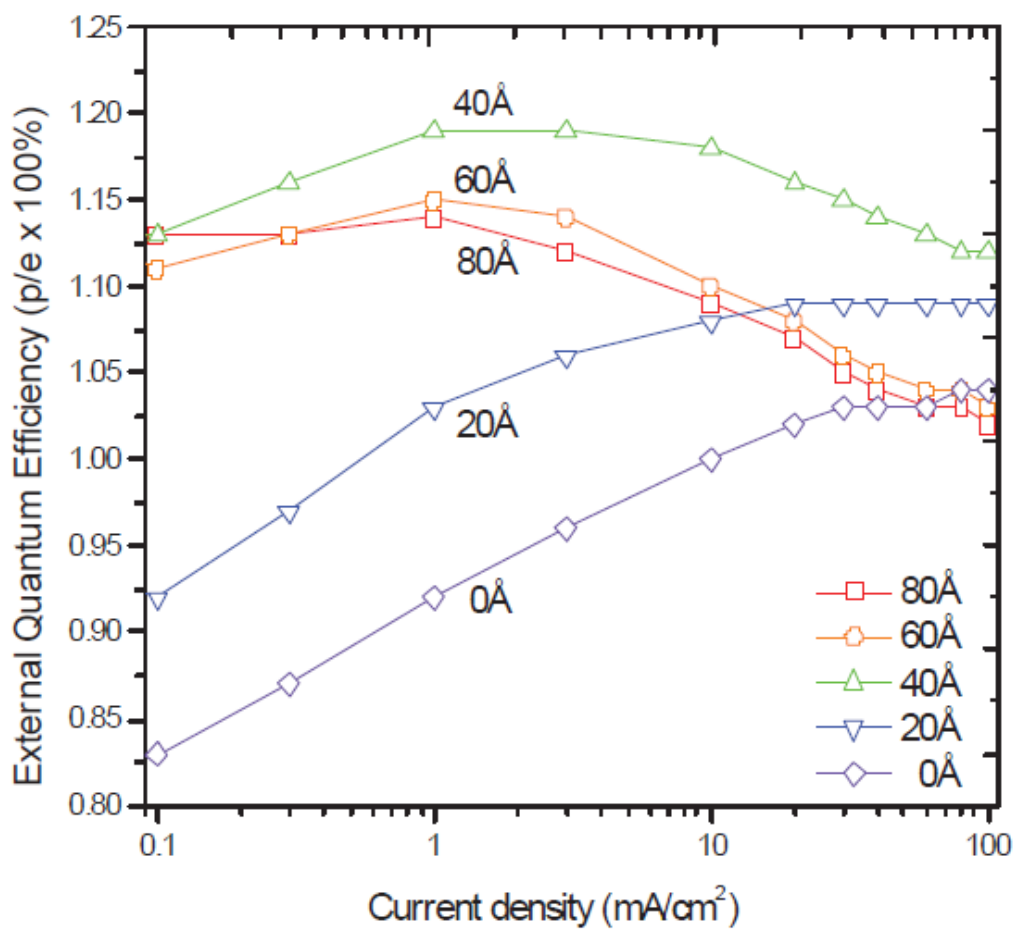

**Figure 3:** Same as Fig.2 using a light-emitting layer diluted with BCP in an ITO/HAT-CN/NPB/BCP:Alq(1:1,dÅ)/Alq/BCP/LiF/Al OLED.